\documentclass
[aps,prl,amsmath,amsfonts,amssymb,twocolumn,superscriptaddress,showpacs]{revtex4}%
\usepackage{times}
\usepackage{amsfonts}
\usepackage{amsmath}
\usepackage{amssymb}
\usepackage{graphicx}%
\providecommand{\U}[1]{\protect\rule{.1in}{.1in}}
\def\ket#1{\mathinner{|{#1}\rangle}}

\begin{document}

%\title{Quantum correlations between qubit and its environment in terms of interatomic distance}
\title{Quantum correlations between each qubit in a two-atom system \\ and the environment in terms of interatomic distance}

\author{K.\ Berrada} 
 %\email{xxx}
\affiliation{The Abdus Salam International Centre for Theoretical Physics, Strada Costiera 11, Miramare-Trieste, Italy}
\affiliation{Laboratoire de Physique Th\'{e}orique, Facult\'{e} des Sciences, Universit\'{e} Mohammed V-Agdal, Av. Ibn Battouta, B.P. 1014, Agdal Rabat, Morocco}
\author{F.\ F.\ Fanchini}
\affiliation{	Departamento de F\'{i}sica, Faculdade de Ci\^{e}ncias, UNESP, Bauru, SP, CEP 17033-360, Brazil}  
\author{S.\ Abdel-Khalek}
\affiliation{Mathematics Department, Faculty of Science, Sohag University, 82524 Sohag, Egypt}
\affiliation{Mathematics Department, Faculty of Science, Taif University, Taif, Saudi Arabia} 
\begin{abstract}
The quantum correlations between a qubit and its environment are described quantitatively in terms of interatomic distance. Specifically, considering a realistic system of two two-level atoms and taking into account the dipole-dipole interaction and collective damping, the  quantum entanglement and quantum discord are investigated, during the dissipative process, as a function of the interatomic distance. For atoms that are initially maximally entangled, it turns out that there is a critical distance where each atom is maximally quantum correlated with its environment. Counterintuitively, the approach of the two atoms can maximize the entanglement between each one and the  environment and, even at the same distance, minimize the loss of entanglement between the pair.
\end{abstract}

\date{\today}
 \maketitle

\section{Introduction}
It is well known that, besides entanglement, other quantum correlations are crucial elements of quantum information theory.
In this plethora of new quantum correlation measures, quantum discord (QD) has emerged as the most frequently used. The QD obeys a monogamic relation with
the entanglement of formation (EOF) and, for a tripartite pure state, QD and EOF obey a conservative relation \cite{law, winter}. Furthermore, QD is connected with the entanglement irreversibility \cite{cornelio}, with entanglement in a measurement \cite{bruss}, and with the entanglement distribution \cite{cavalcanti}.
The dynamics of quantum entanglement and discord in open quantum systems have been widely investigated
in the literature \cite{open,serra}, but few of these studies focus on the way that the system gets entangled with the environment \cite{retamal,serra}. Despite the fact that quantum correlation (the EOF or the QD) does not obey a monogamous equation \cite{ckw}, certainly the way in which each part of a bipartite system becomes quantum correlated with the environment governs the way the entanglement and discord vanish from this system.

This leaves the question, how does a quantum system get quantum correlated with its environment?
Here, we study this problem in a realistic situation, taking two two-level atoms as the qubits.  We take into account
the dipole-dipole interaction, the collective damping and, more importantly, the interatomic distance between them. We investigate how the interatomic distance influences the way that one of these atoms becomes quantum correlated with the environment  and, as we will show, this separation emerges as a crucial variable. We demonstrate that there is an intermediate critical distance, $r_{\rm critical}$, that maximizes the entanglement between the environment and each atom. More importantly, we show that the distance at which each atom gets most entangled with the environment can be exactly the distance where the loss of entanglement between the pair of atoms is at a minimum. This counterintuitive example reveals an important new facet in the construction of a scalable quantum computer, namely that the approach of the qubits can maximize the quantum correlation (EOF and QD) with an environment and, even so, minimize the loss of quantum correlation between the pair of atoms.

This article is organized as follows. In Sec. II, we introduce
some measures of nonclassicality and, in particular,
nonclassical correlations: quantum mutual information,
entanglement and quantum discord. In Sec. III, we present the
model for our system and describe the dependence of the dynamics on the interatomic distance. In Sec. IV,
we study the correlations between one atom and the environment in terms of the interatomic distance using,
the monogamic relation between discord and entanglement in a tripartite system.
We conclude our work in Sec. V.

\section{Quantum correlations}
In this article we use two kinds of quantum correlation to analyze our results: EOF and QD. The EOF is broadly accepted to be necessary for a set of important tasks in
quantum information theory, such  as quantum teleportation \cite{tele}, quantum key distribution \cite{key} and many others. On the other hand, QD emerges as a fundamental quantity of quantum information and includes other kinds of quantum correlations than entanglement \cite{interest,interest1,interest2,interest3,interest4}. %, whose interest of the scientific community increase significantly  %Very recently QD was connected with important tasks as the entanglement irreversibility, the state merging protocol and many other aspects of the quantum information theory.

The EOF is a measure of entanglement, developed about fifteen years ago, with a clear operational interpretation \cite{divi}. For two qubits, an
analytical solution was developed by Wootters in terms of concurrence \cite{concurrence}. In this case, it can be calculated as
\begin{equation}
E(\rho)=H\left({1+\sqrt{1-C^2(\rho)}\over2}\right)
\end{equation}
where $H$ is the binary entropy function defined as
\begin{equation}
H(x)=-x\log_2 x-(1-x)\log_2(1-x),
\end{equation}
and the concurrence is given by
\begin{equation}
C(\rho)=\max\{0,\sqrt{\lambda_1}-\sqrt{\lambda_2}-\sqrt{\lambda_3}-\sqrt{\lambda_4}\},
\end{equation}
where $\lambda_i$ are the eigenvalues, listed in decreasing order,  of $\rho\tilde{\rho}$. $\tilde{\rho}$
is the time-reversed density operator,
\begin{equation}
\tilde{\rho}=(\sigma_y\otimes\sigma_y)\rho^*(\sigma_y\otimes\sigma_y),
\end{equation}
where $\rho^*$ is the conjugate of $\rho$ in the standard basis of two qubits and $\sigma_y$ is the Pauli $y$ operator.

The QD, on the other hand, was originally defined as the mutual information minus the classical correlation \cite{zurek}, where the latter is given by the well-known Henderson and Vedral definition \cite{hen}. Hence,
\begin{equation}
\delta_{AB}^\leftarrow=\mathcal{I}(\rho_{AB})-\max_{\{\Pi_k\}}\mathcal{I}(\rho_A|\{\Pi_k\})
\end{equation}
where $\mathcal{I}(\rho_{AB})$ is the quantum mutual information, which comprehends the total amount of correlation, both classical and
quantum, in a given bipartite quantum state and $\max_{\{\Pi_k\}}\mathcal{I}(\rho_A|\{\Pi_k\})$ is the maximal
classical mutual information when, in this case, a measurement is performed on subsystem $B$ \cite{zurek,hen}. The maximization is carried out over all possible positive operator valued
measures and, in general, it is very hard to carry out, except in some particular cases. If $\rho_A$ ($\rho_B$) is the reduced density operator
of part $A$ ($B$), then the quantum mutual information
is defined as
\begin{equation}
\mathcal{I}(\rho_{AB})=S(\rho_A)+S(\rho_B)-S(\rho_{AB})
\end{equation}
where $S(\cdot)$ is the von Neumann entropy.
The measurement-based mutual information is
often called classical correlation and is given by:
\begin{eqnarray}
\nonumber\mathcal{C}_{AB}^\leftarrow &=&\max_{\{\Pi_k\}}\mathcal{I}(\rho_A|\{\Pi_k\})\\
&=& S(\rho_A)-\min_{\{\Pi_k\}}\sum_k p_k S(\rho_A|\{\Pi_k\}),
\end{eqnarray}
where, $\rho_A|\{\Pi_k\}={\rm Tr}_B(\Pi_k \rho_{AB} \Pi_k)/{\rm Tr}_{AB}(\Pi_k\rho_{AB} \Pi_k)$ is the reduced state of $A$ after obtaining the outcome $k$ in $B$ and $\Pi_k$ is a complete set of positive operator valued
measures that result in the outcome $k$ with probability $p_k={\rm Tr}_{AB}(\Pi_k\rho_{AB} \Pi_k)$.
Finally, the QD is thus defined in terms of the mismatch
\begin{equation}
\delta_{AB}^\leftarrow=\mathcal{I}(\rho_{AB})-\mathcal{C}^\leftarrow_{AB}.
\end{equation}
This definition is in general asymmetric with respect to
the interchange of the subsystems %. Analogously, one is led to define the $\delta_{AB}^\leftarrow$  through
%the entropy of conditional states of system $B$.
%The QD
and it is always a non-negative quantity. Actually, a quantum state has zero discord if and only if there exist a complete
orthonormal basis ${|l\rangle}$ for subsystem $A$ and
some density operator $\rho_B$ for subsystem $B$, such that
$\rho=\sum_lp_l|l\rangle\langle l|\otimes\rho_B$,
 where $|l\rangle$ is an orthonormal set,
${p_l}$ is a probability distribution and $\rho_B$ are quantum
states. %Unfortunately, zero-discord states are of zero
%measure \cite{ferraro} and nonzero values of the quantum discord are
%notoriously difficult to compute because of the required minimization.

\section{The model}
Here we consider a realistic situation, where two identical two-level atoms are coupled to a quantized electromagnetic field \cite{model,model1}. The two atoms are close enough for the transition dipole moment and collective damping to need to be considered. Furthermore, and more importantly, we analyze the influence of the separation between the qubits, that here is given by the interatomic distance. In the interaction picture, the Hamiltonian can be written as:
\begin{eqnarray}
H=&-&\!\!i\sum_{i=1}^2\sum_{\vec{k}s}\left[   \vec{d}_i\cdot\vec{g}_{\vec{k}s}(\vec{r}_i) a_{\vec{k}s}\left(\sigma_i^+e^{-i(\omega_k - \omega_i)t}\right)\right]\nonumber\\
&-&\!\!i\sum_{i=1}^2\sum_{\vec{k}s}\left[   \vec{d}_i\cdot\vec{g}_{\vec{k}s}(\vec{r}_i) a_{\vec{k}s}\left(\sigma_i^-e^{-i(\omega_k + \omega_i)t}\right)\right] + H.c.,\nonumber\\
\end{eqnarray}
where $i$ counts the two atoms and $\vec{k}s$ the field mode. Here,
$\vec{d}_i$ is the transition dipole moment, $\omega_i$ is the transition frequency, $\sigma_i^+$ ($\sigma_i^-$) is the
raising (lowering) operator and $a_{\vec{k}s}$ is the annihilation operator of the field mode  $\vec{k}s$ . With this field mode, we associate
a vector $\vec{k}$, a frequency $\omega_k$, and an index of polarization $s$. The coupling constant is given by
\begin{equation}
\vec{g}_{\vec{k}s}(\vec{r}_i) =\sqrt{\frac{\omega_k}{2\epsilon_0\hbar V}}\hat{e}_{\vec{k}s}\exp{i\vec{k}\cdot\vec{r}_i},
\end{equation}
where $V$ is the quantization volume, $\epsilon_0$ is the vacuum permittivity, and $\hat{e}_{\vec{k}s}$ is the electric field polarization vector. Finally, $\vec{r}_i$ is the
position of the $i$-th atom.
Here, we suppose that the environment begins in the vacuum state and we assume that the rotating-wave approximation is valid. In this case, considering a Markovian approximation, the dynamics of the two atoms can be written as
\begin{eqnarray}
\frac{\partial\rho}{\partial t}=&-&\omega_0\sum_{i=1}^2[\sigma_i^z,\rho]-\sum_{i\ne j}\Omega_{ij}\left[\sigma_i^+\sigma_j^-,\rho \right]\nonumber\\
&-&\frac{1}{2}\sum_{i,j=1}^2\gamma_{ij}\left(\sigma_i^+\sigma_j^-\rho-2\sigma_j^-\rho\sigma_i^+ \rho\sigma_i^+\sigma_j^-\right)\label{masterequation}
\end{eqnarray}
where $\sigma_i^z$ is the Pauli $z$ operator of the $i$th atom, $\gamma_{ii}\equiv\gamma$ is the spontaneous decay rate, and
$\gamma_{ij}$, $\Omega_{ij}$ describe the collective damping and dipole-dipole interaction. Explicitly, we have,
\begin{eqnarray}
\gamma_{ij}\!&=&\!\frac{3}{2}\gamma\left[1-\left(\vec{d}.\vec{r}_{ij} \right)^2 \right]\frac{\sin\left[ k_0r_{ij}\right]}{k_0r_{ij}}\nonumber\\
&+&\frac{3}{2}\gamma\left[1-3\!\left(\vec{d}.\vec{r}_{ij} \right)^2 \right]\!\!\left[ \frac{\cos\left[ k_0r_{ij}\right]}\nonumber
{(k_0r_{ij})^2} - \frac{\sin\left[ k_0r_{ij}\right]}{(k_0r_{ij})^3} \right]\\
\end{eqnarray}
and
\begin{eqnarray}
\!\!\!\Omega_{ij}\!=\!\!&-&\!\!\!\frac{3}{4}\gamma\left[1-\left(\vec{d}.\vec{r}_{ij} \right)^2 \right]\frac{\cos\left[ k_0r_{ij}\right]}
{k_0r_{ij}}\nonumber\\
&+&\!\!\!\frac{3}{4}\gamma\left[1-3\!\left(\vec{d}.\vec{r}_{ij} \right)^2 \right]\!\!\left[ \frac{\sin\left[ k_0r_{ij}\right]}{(k_0r_{ij})^2}\nonumber
+ \frac{\cos\left[ k_0r_{ij}\right]}{(k_0r_{ij})^3} \right],\\
\end{eqnarray}
where $k_0=\omega_0/c$, $r_{ij}=|r_i-r_j|$ is the distance between the two atoms, $\vec{d}$ is the unit vector along the atomic
transition dipole moment and $\hat{r}_{ij}$ is the unit vector along the interatomic axis.
Note that in Eq. ({\ref{masterequation}}) the electromagnetic field has been formally eliminated, to be replaced by an effective atom-atom interaction. Furthermore, it is important to emphasize that the master equation takes into account for the spontaneous emission process.

To illustrate the dependence of the dynamic on the atomic distance we focus on two important classes of pure states given by
\begin{equation}\label{bell23}
\ket{\Phi} = \alpha|01\rangle + \sqrt{1-\alpha^2}|10\rangle,
\end{equation}
and
\begin{equation}\label{bell14}
\ket{\Psi} = \beta|00\rangle + \sqrt{1-\beta^2}|11\rangle
\end{equation}
where $0$ and $1$ represent the atom two-level system. These initial conditions are of fundamental importance in
quantum information theory since they include the four Bell states that can be used on their own for universal quantum computation and are associated with an unit of entanglement called \textit{e-bit}.
It should be understood that, despite this limitation of the initial state, our analysis is general and the dynamics could be calculated for any initial state.

\section{Quantum correlations between one atom and the environment}
It is important to note that to calculate how one of the atoms quantum correlates with the environment is a simple but not
direct task. To elucidate this aspect we begin by supposing just one two-level atom is interacting with an environment. If the whole
system (atom plus environment) begins as a pure state, the manner in which this atom gets entangled with the environment can be calculated
in a simple manner: the whole system is pure and the von Neumann entropy of the atom is the entanglement of formation. Now, extending
the analysis to our situation, where two initially pure atoms interact with an environment, how could we calculate the entanglement between
one of the atoms and its environment? In this case, since the bipartite system composed of one atom plus its environment is a mixed
state, the von Neumann entropy is not the right answer. For this purpose we use the monogamic relation between the EOF and the QD \cite{law}.

We suppose the initial state of the environment as the vacuum state since the whole system (two atoms plus environment)
is pure. In this case, the monogamic relation gives:
\begin{equation}
E_{AE}=\delta_{AB}^\leftarrow+S_{A|B}
\end{equation}
and
\begin{equation}
\delta_{AE}^\leftarrow=E_{AB}+S_{A|B}.
\end{equation}
Thus, as we observe, the entanglement of formation and the quantum discord between the atoms can be used to calculate the
quantum correlations between the environment and one of the atoms. Curiously, for this task we do not need any information about the state of the environment since it is completely traced out to calculate the dynamics. Actually, for a tripartite pure state it is always possible to infer the EOF and the QD if just one of the possible bipartitions is known. Moreover, the quantum discord between $A$ and $E$ can be calculated analytically without any approximation \cite{LII}.

To explore the influence of interatomic distance on the behavior of the quantum correlations, we
plot, for various values of scaled time $\gamma t$, the correlations between different bipartitions as a function of
distance $k_0 r$. We emphasize that the choice of the
range of values for the plot axes is made only for clarity and that other values of the parameters
$\alpha$ and $\beta$ in the initial states do not alter the critical distance at which one of the atoms is maximally quantum correlated with the environment.
\begin{figure}[htbp]
\begin{center}
\includegraphics[width=.45\textwidth]{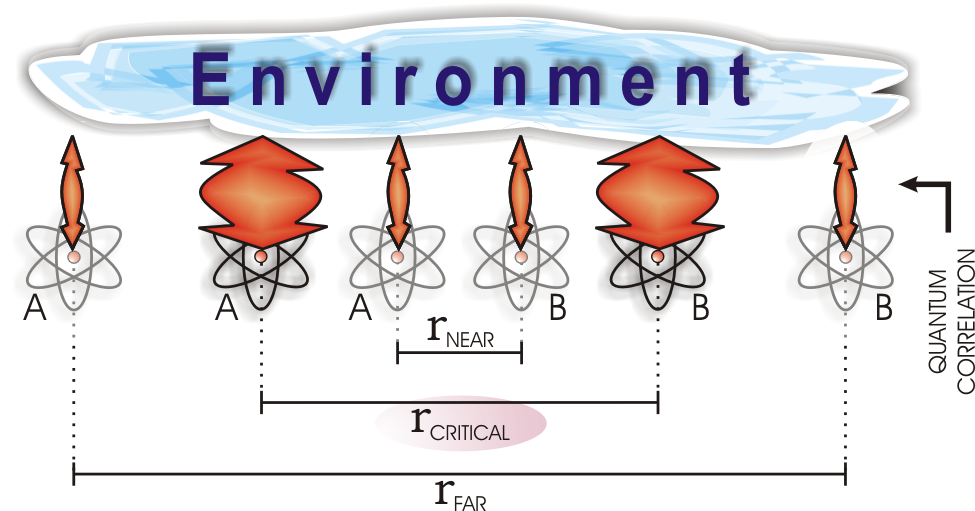} {}
\end{center}
\caption{(Color Online) An illustrative scheme to elucidate the critical distance where each atom is maximally quantum correlated
with the environment.}\label{fig001}
\end{figure}

\begin{figure}[htbp]
\begin{center}
\includegraphics[width=.50\textwidth]{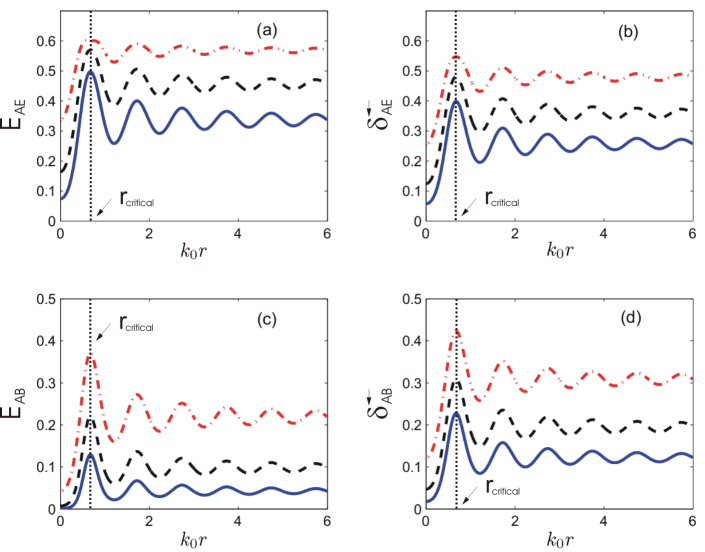} {}
\end{center}
\caption{(Color Online) (a) $E_{AE}$, (b) $\delta_{AE}^\leftarrow$, (c) $E_{AB}$, and (d) $\delta_{AB}^\leftarrow$ as a function of
the interatomic distance for various values of the scaled time. The dashed-dotted (red) line is for $\gamma t=1$, the dashed (black) line is for
$\gamma t=1.5$ and the solid (blue) line is for $\gamma t=2$.   The two-qubit initial state is given by Eq. \eqref{bell23} with $ \alpha^2=1/2$.}
\end{figure}

\begin{figure}[htbp]
\begin{center}
\includegraphics[width=.50\textwidth]{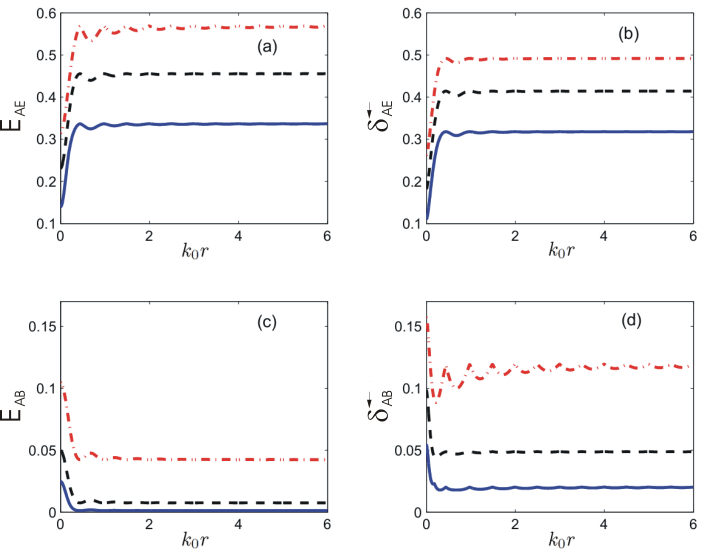} {}
\end{center}
\caption{(Color Online) (a) $E_{AE}$, (b) $\delta_{AE}^\leftarrow$, (c) $E_{AB}$, and (d) $\delta_{AB}^\leftarrow$ as a function of
the interatomic distance for various values of the scaled time. The dashed-dotted (red) line is for $\gamma t=1$, the dashed (black) line is for
$\gamma t=1.5$ and the solid (blue) line is for $\gamma t=2$.   The two-qubit initial state is given by Eq. \eqref{bell14} with $ \beta^2=1/2$.}
\end{figure}

In Figs. 2 and 3, we show the variation of correlations $E_{AE}$, $\delta_{AE}^\leftarrow$, $E_{AB}$, and $\delta_{AB}^\leftarrow$ with
interatomic distance for various values of the scaled
time, and for the initial Bell states \eqref{bell23} and \eqref{bell14}, respectively. The dashed-dotted
line is for $\gamma t=1$, the dashed line is $\gamma t=1.5$, and the solid line is for $\gamma t=2$.

From Fig. (2), we note that for each partition, $AE$
and $AB$,
the correlations EOF and QD behave similarly with respect to distance $k_0 r$.
Interestingly, we find that for each instant, there is one maximum of
the correlations $E_{AE}$ and $\delta_{AE}^\leftarrow$
at a specific value of the interatomic distance, $r_{\rm critical}$. Hence, in a short time, one especial distance maximizes the EOF and QD between each qubit and the environment. Furthermore, for the initial condition given by Eq.  \eqref{bell23}, the critical distance that maximizes the quantum correlation with the environment is exactly the one that least disturbs the quantum correlations between the atoms.
%\textcolor{red}{ and shows that minimizes the quantum correlations with the environment is far to be a good strategy to minimize the loss of quantum correlations between the pair.}
{This is a very counterintuitive feature since, in this case, diminishing the quantum correlations between each atom and the environment is \textit{not} a good strategy to preserve the quantum correlations between the atoms.}
This means that the maximal correlations between one atom and the environment are very sensitive to the distance $r$ and that
the decay rate of correlations between one subsystem and its environment may be controlled
through the separation between the subsystems. As we see in Fig. 2(c), for a fixed time ($\gamma t=1$), the entanglement between the atoms can vary by about $0.2$ as the interatomic distance rises from $r_{critical}$. For a separation of about $k_0 r=0.7$, the environment reduces the initial entanglement to approximately $0.4$, while at $k_0 r=1$, the entanglement goes down to less than $0.2$. Furthermore, the relation between the distance and the loss of entanglement is not monotonic, since at a distance of about $k_0 r=1.8$ the environment reduces the initial entanglement to about $0.3$. These results make the distance between the qubits a relevant parameter to taken into account in the construction of a working quantum computer.

On the other hand, it is important to note that the behavior of the correlations, $E_{AB}$ and $\delta_{AB}^\leftarrow$, present a different response for a different initial state. For example, from Figs. 3(c) and 3(d), we find that the increase of EOF and QD with the environment is accompanied by a decrease of the correlations between the subsystems. However, it is important to emphasize that even in this case a critical distance exists, where each atom entangles fast with the environment.

\section{Conclusion}
To summarize, we have studied the distribution of the quantum correlations, given by the EOF and the QD, between the qubits and the environment. We focused on the way that the distance between two qubits, given by two atoms, can affect the manner in which each one becomes quantum correlated with the surroundings. In our analysis a critical distance emerged, where the quantum correlation of each atom with the environment is at a maximum. Counterintuitively, we showed that this critical distance can be the one that minimizes the loss of quantum correlations between the pair of atoms. Indeed, the gap between the qubits emerged as a fundamental element in the construction of a scalable quantum computer, and the size of this gap was shown to be an important variable to take into account.
The present results also suggest that in a future analysis, initially mixed states should be considered since the influence of finite-temperature environments on the critical distance could then be contemplated.

This work was partially supported by FAPESP and CNPq through the National Institute for Science and Technology of Quantum Information (INCT-IQ).
%The authors acknowledge the hospitality of the Abdus
%Salam International Centre for Theoretical Physics (Trieste, Italy).

\end{document}